# Closed Form Expressions of the Nonlinear Interference for UWB Systems

Pierluigi Poggiolini[(1)], Mahdi Ranjbar-Zefreh[(2)]

[(1)] DET, Politecnico di Torino, Torino, Italy, pierluigi.poggiolini@polito.it
[(2)] CISCO Systems S.R.L., Vimercate (MB), Italy

**Abstract** *We present a comprehensive closed-form GN/EGN model supporting ultra-wide-band systems spanning 50 THz of optical bandwidth. We show a case-study of 10x100km of SMF where we gradually increase the number of channels across the C,L,S,U,E bands while optimizing launch power.*

**Introduction**
Many technologies are currently competing in the quest for increasing the throughput of optical links. They can be broadly classified as either "space-division-multiplexing" (SDM) or "ultra-wide-band" (UWB). All SDM and some UWB technologies require that new cables be deployed. The notable exception is UWB over standard single-mode fiber (SMF). This circumstance makes UWB over SMF very attractive for carriers and is spawning substantial interest in the research and industrial community.

UWB over SMF consists of extending the transmission bandwidth beyond the C band. The extension to L-band, which is already commercially available, can be considered a first success of this principle. Research is now focusing on the other SMF bands, primarily S and O but also E and U.

These bands present challenges that are both technological and propagation-related. On the technology side, suitable amplification is possibly the most significant. Regarding propagation, inter-channel stimulated Raman scattering (ISRS) and higher loss are the primary ones, but also low dispersion is significant, especially in the E and O bands. To study the potential of UWB, suitable propagation models are needed, capable of being very fast and still accurate over such a large spectral extension. All effects, including Kerr, present substantial variations across bands, which need to be accounted for.

The need for fast evaluation stems from the necessity of performing complex optimization, for instance of per-channel power, to either maximize total throughput or to achieve other system-related requirements. Fast evaluation makes it difficult to use the integral form of broadly used models such as the GN or EGN. Approximate NLI Closed-Form Models (CFMs) derived from GN/EGN have been recently developed by several groups, among which UCL [1]-[6] and Politecnico di Torino (PoliTo) [7]-[12]. The resulting CFMs are roughly equivalent, though significant differences exist. In this paper, the PoliTo CFM approach [7]-[12] is addressed and extended to UWB. As case-study, consisting of the estimation of the throughput of a UWB SMF system across several bands, is presented, with channel power optimization. Note that other interesting GN-model UWB extensions have been proposed as well [13]-[17].

**The UWB Propagation Model**
The CFM we describe approximates the GN/EGN models. It is based on the CFM proposed in [7], extended to handle ISRS and frequency-dependent loss and dispersion [8], [9]. Its accuracy was then improved using a machine-learning approach [10]. To support UWB, we include frequency-dependent Kerr and Raman, as well as higher-order dispersion through $\beta_4$.

A key staple of this CFM is that channel loss, in a given span, is described as [8]:

$$\alpha_n(z) = \alpha_{0,n} + \alpha_{1,n} \cdot \exp(-\sigma_n \cdot z) \quad (1)$$

where the index $n$ identifies the channel located at frequency $f_n$. Eq.(1) is loosely inspired by a physical picture where $\alpha_{0,n}$ is fiber loss in the absence of ISRS; $\alpha_{1,n}$ is the change in loss due to ISRS at the start of the span; $\sigma_n$ is related to how fast ISRS vanishes along the span as overall optical power goes down. Once $\alpha_{0,n}, \alpha_{1,n}, \sigma_n$ have been assigned to each channel, then the NLI calculation is fully closed-form (CF). However, the assignment of $\alpha_{0,n}, \alpha_{1,n}, \sigma_n$ is not fully-CF, since we deemed that a fully-CF approach to it might lead to excessive error. This is the only non-CF step in the procedure and is done as follows. First, we numerically solve the ISRS differential equations (Eq. (1) in [18]) and find the exact power evolution of every channel. Then, Eqs. (30.1) and (30.2) in [9] provide a CF best-fit for $\alpha_{0,n}, \alpha_{1,n}$, for a given $\sigma_n$. Finally, numerically optimizing over $\sigma_n$ provides the overall best-fit for $\alpha_{0,n}, \alpha_{1,n}$ and $\sigma_n$. Note that for these calculations the Raman gain coefficient $C_R$ is needed. $C_R$ depends on both the "pump channel" frequency $f_p$ and the difference between pump and signal channels $\Delta \nu = f_p - f$, that is: $C_r(f_p, \Delta \nu)$. See later

$$P_n^{\text{NLI}} = \frac{16}{27} P_n \sum_{\substack{1\leq m\leq N_c, 0\leq j\leq 1 \\ 0\leq k\leq M, 0\leq q\leq M}} \frac{\gamma_{n,m}^2 \cdot P_m^2 \cdot \rho_m \cdot (2-\delta_{m,n}) \cdot e^{-4\alpha_{1,m}/\sigma_m} \cdot (-1)^j}{2\pi R_m^2 \cdot k!\, q!\cdot \bar{\beta}_{2,m} \cdot (4\alpha_{0,m} + (k+q)\sigma_m)} \cdot \left(\frac{2\alpha_{1,m}}{\sigma_m}\right)^{k+q} \psi_{m,n,j,k} \quad (2)$$

for more details on $C_r$. Then, the NLI power $P_n^{\text{NLI}}$ generated in a given span, affecting the $n$-th channel, can be found using Eq.(2) above, where: $P_m$, $R_m$ and $\bar{\beta}_{2,m}$ are for the $m$-th channel, respectively, the launch power into the span, the symbol rate and the '*effective dispersion*'. The latter is given by:

$$\bar{\beta}_{2,m} = \beta_2 + \pi\beta_3(f_n + f_m - 2f_0) + \frac{2\pi^2}{3} \cdot \beta_4[(f_n - f_0)^2 + (f_n - f_0)(f_m - f_0) + (f_m - f_0)^2] \quad (3)$$

where $f_0$ is the frequency where $\beta_2$, $\beta_3$ and $\beta_4$ are measured. Note that $\beta_4$ is necessary to correctly represent dispersion in UWB. The non-linearity coefficient $\gamma_{n,m}$ has the UWB expression [19]:

$$\gamma_{n,m} = \frac{2\pi f_n}{c} \frac{2n_2}{A_{\text{eff}}(f_n) + A_{\text{eff}}(f_m)} \quad (4)$$

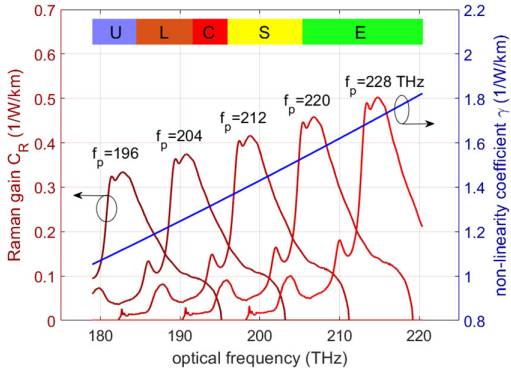

**Fig.1:** Blue curve: non-linearity coefficient $\gamma$ vs. frequency. Red curves: Raman gain coefficient $C_r$ for a few different values of the "pump channel" frequency $f_p$. Top: freq. bands.

where $n_2$ is the nonlinear (Kerr) refractive index. An example of $\gamma_{n,n}$, is shown in Fig. 1 for a SMF, as a function frequency. Note that $\gamma_{n,n}$ weighs SPM (or SCI), whereas $\gamma_{n,m}$ with $n \neq m$ weighs XPM (or XCI). We neglect here the effect described in [5]. Furthermore, in Eq.(2):

$$\psi_{m,n,j,k} = \text{asinh}\left(\frac{\pi^2 \bar{\beta}_{2,m} R_n \cdot (f_m - f_n + (-1)^j \cdot R_m/2)}{2\alpha_{0,m} + k\sigma_m}\right) \quad (5)$$

Also in Eq.(2), $\delta_{n,m}$ is 1 if $n = m$ and 0 otherwise, and $N_c$ is the number of WDM channels. $M$ represents the order of a series expansion, whereby more terms are called into play as the strength of ISRS requires, and is set by [9]:

$$M = \max(\lfloor 10 \cdot |2\alpha_{1,n}/\sigma_n|\rfloor + 1) \quad (6)$$

where the maximum is calculated across all channels ($1 \leq n \leq N_c$) in the WDM comb.

Finally, $\rho_m$ is the machine-learning-based 'correction term' given by Eq.(14) (fed by Table IV) in [10]. Note that $\rho_m$ also depends on the span index, like essentially all other quantities in Eq.(2). However, to avoid clutter, we have omitted to indicate such dependence. $\rho_m$ improves the accuracy of the CFM Eq.(2), as shown in [10], allowing it to closely approach the EGN model. $\rho_m$ depends in closed-form on $R_m$, on the dispersion accumulated till the considered span and on the modulation format of the $m$-th channel (see [10] for details).

As mentioned, Eq.(2) provides the $P_n^{\text{NLI}}$ generated on the $n$-th channel within one span of the link. To obtain the total NLI at the end of the link it is enough to sum the $P_n^{\text{NLI}}$ of each span. The total NLI can then be used together with ASE noise to find the generalized OSNR for the $n$-th channel at the end of the link.

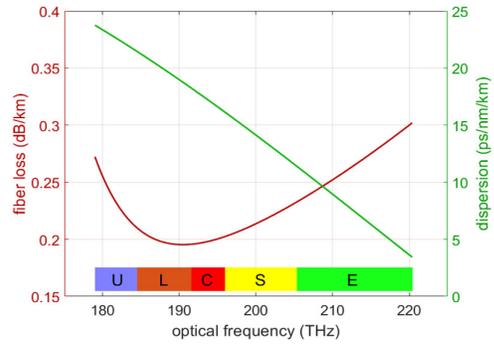

**Fig.2:** Loss (red line) and dispersion (green line) vs.

**Case-Study: Throughput of a UWB System**
To test the effectiveness of the model, we use it to assess the total throughput of a system based on 10 spans of 100km of SMF, populating progressively the C,L,S,U and E bands. The assumed fiber loss (with "zero water-peak") and dispersion are shown in Fig.2. The NL coefficient $\gamma_{n,m}$ is given by Eq.(4) assuming $n_2$=2.6e-20, and $A_{\text{eff}}(f)$ is given by the empirical formula Eq.(8) in [20] with numerical aperture NA=0.124 and core radius 4.1$\mu m$. The Raman gain coefficient $C_r(f_p, \Delta v)$ was experimentally characterized on a SMF and then scaled in frequency according to Eqs.(37)-(39) in [21]. Fig. 1 shows the plot of a few instances of $C_r(f_p, \Delta v)$ for different pump frequencies. The link was assumed to use lumped amplifiers with the following noise figures: C 5dB, L 6dB, S and E 7dB, U 8dB.

The transmitted WDM signal consisted of 64GBaud channels spaced 75 GHz. We did not impose guard intervals between bands for the sake of simplicity. To assess system throughput realistically, it is essential to assume a net transceiver user information rate (IR). We adopted the IR shown in Fig. 3, which is expected to be provided by the next generation of commercial transceivers.

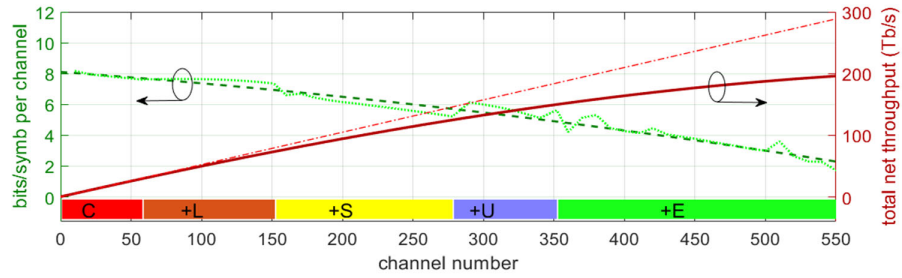

**Fig.5:** Red solid line: total system throughput vs. number of WDM channels at the end of a 10x100km SMF link. Channels are progressively added in the bands shown by the band indicator at the bottom of the figure. Red dashed line: linear extrapolation with same slope as for the C band. Green dots: bits/symbol carried by the next added channels; dashes: smoothed version.

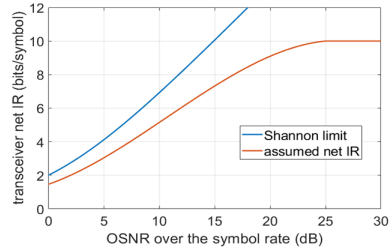

**Fig.3:** Assumed transceiver net information rate vs.

We started populating the WDM spectrum at the low-frequency end of the C-band and then went on in 10-channel increments, filling up first the C, then the L, S, U and E band. Launch power was optimized as a cubic polynomial curve, independently in each band, identical at the start of each span. Fig.4a shows the obtained net IR per channel as a function of frequency, for different sets of occupied bands. Fig.4b shows the corresponding optimum launch power profile. The plots clearly indicate that the C and L band operate at high IR (about 7.5 bits/symb) in all configurations. The S band is, on average, more than 2 bits/symb lower than C+L. The optimum power distribution clearly shows that in the C+L+S case the S-band acts as a distributed Raman pump for C+L channels, whose optimum launch power is several dBm below the case of only C+L. The S-band improves when the E-band is present because the latter Raman-pumps the S-band channels. However, E-band channels are poorly performing. Fig.5 shows total throughput vs. channel number, and next-channel IR contribution. From the plot it appears that S-band provides extra throughput at reasonable efficiency while the E band appears to be much less attractive. These results are in general agreement with very recent similar studies [6],[16],[17]. Note that we tried adding the O-band but the extra throughput was very small.

**Comments and Conclusion**

The availability of fast and reliable closed-form-models (CFMs) allows to perform insightful investigations and optimizations of UWB systems. We presented a comprehensive CFM, based on the GN/EGN model, accounting for UWB-frequency-dependent: Kerr and Raman effects, loss and dispersion.

We showed a case-study of UWB over 10x100km SMF, across the C,L,S,U and E bands, intended to illustrate the effectiveness of the CFM. In the analysis we also included realistic transceiver performance. The results show very diversified launch power optimization profiles depending on how many bands are populated. They also show that the S-band, despite degraded propagation features, may ideally provide a 60%-65% throughput increase vs. C+L. In our opinion this case-study shows that the proposed CFM is a powerful tool to explore and optimize the next generation UWB systems.

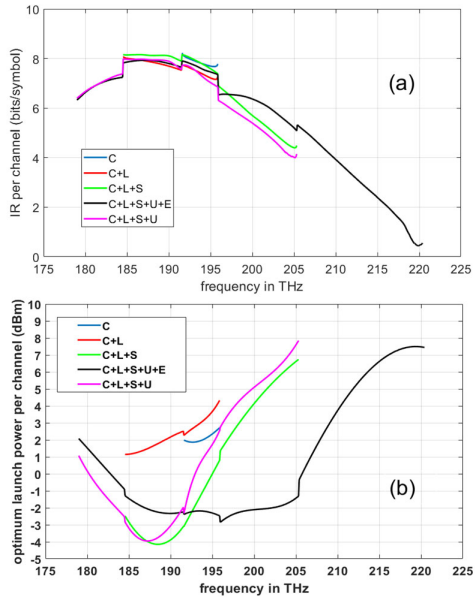

**Fig.4:** **(a)** Obtained Information rate per channel at the end of the test 10x100km SMF link and **(b)** optimum launch power, both (a) and (b) vs. channel frequency, for systems using the different sets of bands as indicated in the legend.

**Acknowledgements:** This work was supported by Cisco Systems through the OPTSYS 2022 research contract with Politecnico di Torino and by the PhotoNext Center of Politecnico di Torino. The Authors would like to thank Fabrizio Forghieri and Stefano Piciaccia from CISCO Photonics for the fruitful discussions.